\documentclass[floats,preprint,prl,aps]{revtex4}
\usepackage{latexsym}
\usepackage{graphicx}
\newcommand\beq{\begin{equation}}
\newcommand\eeq{\end{equation}}
\newcommand\bea{\begin{eqnarray}}
\newcommand\eea{\end{eqnarray}}
\newcommand\non{\nonumber}
\newcommand\bib{\bibitem}

\begin{document}
\title{\bf Quantum phase transition of light 
in coupled  
optical cavity arrays: A renormalization group study}

\author{\bf Sujit Sarkar}
\address{\it Poornaprajna Institute of Scientific Research,
4 Sadashivanagar, Bangalore 5600 80, India.\\
}
\date{\today}

\begin{abstract}
We study the quantum phase transition of light of a system when atom trapped in
microcavities and interacting through the exchange of virtual photons. We predict
the quantum phase transition between the photonic Coulomb blocked induce
insulating phase and anisotropic exchange induced photonic superfluid phase 
in the system due to the existence of two Rabi frequency
oscillations. 
The
renormalization group equation shows explicitly that for this system there is
no self-duality. The system also shows two Berezinskii-Kosterlitz-Thouless (BKT) transitions for
the different physical situation of the system.
The presence of single Rabi frequency oscillation in the system leads to
the BKT transition where system shows  
the quantum phase transition from 
photonic metallic state to the Coulomb blocked induced 
insulating phase. 
For the other BKT transition when the z-component of exchange interaction is absent,
the system shows the transition from the photonic metallic state
to the photonic superfluid phase. We also predict the commensurate to incommensurate
transition under the laser field detuning. 
\\

Pacs: 42.50.Dv, 42.50.Pq, 03.67.Bg, 75.10.Jm 
\end{abstract}
\maketitle

{\bf Introduction:} 

The recent experimental success in engineering strong interaction
between the photons and atoms in high quality micro-cavities opens up
the possibility to use light matter system as quantum simulators for
many body physics [1-18].
The authors of Ref. (\cite{hart1},\cite{hart2}, \cite{suj2}) have shown that
effective spin lattice can be generated with individual atom in the
microcavities that are coupled to each other via exchange of virtual
photons. The two states of spin polarization are represented by the
two long lived atomic levels in the system.\\
A Many body Hamiltonians can be created and probed in coupled cavity arrays.
In our previous study, we have explained explicitly the basic physics
of the formation of micro-optical cavity \cite{suj2}.
Atoms in the cavity are used for detection and also for generation of
interaction between photons in the same cavity.
This artificial system can act as a quantum simulator [1-6].
This micro-cavity system shows the different quantum phases and
quantum phase transition (QPT) of photonic states of the system.\\
QPT associate with the fundamental changes that occurs
in the macroscopic nature of the matter at zero temperature due to
the variation of some external parameter. Quantum phase transitions
are characterized by the drastic change in the ground state
properties of the system driven by the quantum fluctuations \cite{subir,indrani}. \\
In this paper, we derive the renormalization group equation
for the continuum field theoretical study of the model Hamiltonian.
We predict with the existence of photonic metallic state,
the anisotropic exchange induces the photonic superfluid phase and photon
blocked induces the insulating phase. The transition from
the photonic superfluid phase to the photon blocked induced insulating
phase is the second order quantum phase transition. Whereas the
transition from photonic metallic state to the photonic insulating
state is the Berezinski-Kosterlitz-Thouless (BKT) 
transition \cite{subir,gia,druf} where the anisotropic exchange 
interaction is absent.
The other BKT transition occurs when we consider the transition 
from photonic metallic state to the photonic superfluid state 
when the z-component interaction is absent. 
\\
The quantum state of engineering of the cavity QED system is
in the state of art due to the
rapid technological development in this
field [1]. Therefore one can achieve the all states
through the proper manipulation of laser frequencies, detuning
field, Rabi frequency oscillations.
In our study, the photonic metallic state is the Luttinger liquid
phase of the microcavities array system. The photonic superfluid
phase is originated from the anisotropic exchange interactions which
discuss in Ref.\cite{jor}. To the best of our knowledge, 
the explicit study of the different quantum phases through
the RG equations and
the proposal
of photonic metallic state is the first in the literature for
the Cavity QED system. Now we discuss in detail about the photon 
blocked induce insulating phase. The photon blocked is
a dressed state of strongly coupled atom-cavity system. In
this phase the inhivitation of resonant absorption of a second
photon if one photon is already resonantly absorbed. This
phenomena was first observed by Birnbaum $et~al.$ \cite{birn}. In this
studies they have predicted the photon blocked in the
single atom in the cavity \cite{birn}. Here we predict this phenomena 
for the array of cavities, where the atoms in different cavities interacting
through virtual photon emission/absorption process.\\ 
We will see after the continuum field theoretical study 
that our model Hamiltonian contains two 
strongly relevant and mutually nonlocal perturbation over
the Gaussian (critical) theory.
In such a situation the strong coupling fixed point is usually
determined by the most relevant perturbation whose amplitude
grows up according to its Gaussian scaling dimensions and
it is not much affected by the less relevant coupling terms.
However, this is not the general rule if the two operators
exclude each other, i.e., if the field configurations which
minimize one perturbation term do not minimize the other.
In this case interplay between the two competing relevant
operators can produce a novel quantum phase transition through
a critical point or a critical line. Therefore, we would like to
study the RG equation to interpret the quantum phases of the system.\\
{\bf {\large Model Setup and Motivation } } \\
We would like to solve the quantum phases and phase boundaries
of coupled cavity arrays by establishing the quantum spin chain
model of the system \cite{hart1,hart2,suj2}. 
At first we discuss the relation between the spin operators and
the atom-photon system.
Our starting point is the Janes-Cummings Hamiltonian,
$ H~ = g ( {\sigma} a^{\dagger} + h.c ) $.
Cavity
mode represent by the bosonic operators ($a, {a}^{\dagger}$ ) and atomic
mode represent by the fermionic operator (${\sigma}, {\sigma}^{\dagger}$).
Where $\sigma$ is the Pauli matrix which transform one excitation from
the radiation field to the atomic field. Therefore, one can write
\[ {\sigma}^{\dagger} = \left (\begin{array}{cc}
      0 & 1 \\
    0  & 0
        \end{array} \right ) \], similarly one
can write for ${\sigma}^{-}$ with $ <1| = (1,0)$ and $ <0|= (0,1) $. The
excitation in this system only transfer between atom and photon in the cavity.
For a fixed number ($n$) of total excitation, one can express
the manifold: ${H_n}= \{ |0,n>, |1,n-1> \}$ provided $n \geq 1 $. Here
$ |0, n> $ and $| 1, n-1 >$ are represent atom in the ground state with
$n$ photon and excited states of the atom with ($n-1$) photon respectively.
We consider the initial state $|e, n-1> $, we obtain the
state $|g,n>$ by the following operation,
$ |g,n> = {\sigma} a^{+} |e, n-1> $. Therefore, we may write the following
relation based on the conservation of the number of excitation.
$ {\sigma}^{\dagger} a |e, n-1> =~0~ = {\sigma} a^{\dagger} |g,n>  $
and $ {\sigma}^{\dagger} a |g, n> = |e, n-1> $. These relations are nothing
but the properties of spin operators acting on the spinors in the z basis.\\
Micro-cavities of a photonic crystal are coupled through the exchange of photons.
Each cavity consists of one atom with three levels in the energy
spectrum, two of them are long lived
and represent two spin states of the system and the other represent excited
states (fig. 1 and fig. 2 of Ref. \cite{hart1,hart2}). Externally
applied laser and cavity modes couple to each atom of the cavity. It may 
induce the Raman transition between these two long lived energy levels. Under a
suitable detuning between the laser and the cavity modes, virtual photons
are created  
in the cavity which mediate interactions with another atom in a 
neighboring cavity. One can eliminate the excited states 
by choosing the appropriate detuning between the applied laser
and cavity modes. 
Then one can achieve only two states per atom in the long 
lived state and the system can be described by a spin-1/2 
Hamiltonian \cite{hart1,hart2}.\\ 
The Hamiltonian of the system consists of three parts:
\beq
H ~= ~ {H_A} ~+~ {H_C}~+~{H_{AC}}
\eeq     
Hamiltonians are the following
\beq
{H_A} ~=~ \sum_{j=1}^{N} { {\omega}_e } |e_j > <e_j | ~+~ 
{\omega}_{ab} |b_j > <b_j | 
\eeq
where $j$ is the cavity index. ${\omega}_{ab} $ and ${\omega}_{e} $ are 
the energies of the state $ | b> $ and the excited state respectively. The
energy level of state $ |a > $ is set as zero. $|a>$ and $|b> $ are
the two stable state of a atom in the cavity and $|e> $ is the
excited state of that atom in the same cavity. 
The following Hamiltonian  describes photons in the cavity,
\beq
 {H_C} ~=~ {{\omega}_C} \sum_{j=1}^{N} {{a_j}}^{\dagger} {a_j} ~+~
{J_C} \sum_{j=1}^{N} ({{a_j}}^{\dagger} {a_{j+1}} + h.c ),  
\eeq 
where ${a_j}^{\dagger}({a_j})$ is the photon
creation (annihilation) operator for the photon field in the $j $'th cavity, ${\omega}_C $
is the energy of photons and $ J_C $ is the tunneling rate of photons
between neighboring cavities.
Interaction between the atoms and photons and also by the driving lasers
are described by
\beq
{H_{AC}}~=~ \sum_{j=1}^{N} [ (\frac{{\Omega}_a}{2} e^{-i {{\omega}_a} t} +
{g_a} {a_j}) |e_j > < a_j | + h.c] + [a \leftrightarrow b ] . 
\eeq
Here ${g_a} $ and ${g_b} $ are the couplings of the cavity mode for the
transition from the energy states $ |a > $ and $ | b> $ to the excited state.
${\Omega}_a $ and ${\Omega}_b $ are the Rabi frequencies of the lasers
with frequencies ${\omega}_a $ and $ {\omega}_b $ respectively.\\
The authors of Ref. \cite{hart1,hart2}
have derived an effective spin model by considering the following physical
processes:
A virtual process regarding emission and absorption of
photons between the two stable  states of neghbior cavity yields the resulting 
effective Hamiltonian as
\beq
{H_{xy}} = \sum_{j=1}^{N}  B {{\sigma}_j}^{z} ~+~\sum_{j=1}^{N} 
(\frac{J_1}{2} {{\sigma}_j}^{\dagger} {{\sigma}_{j+1}}^{-} ~+~
\frac{J_2}{2} {{\sigma}_j}^{-} {{\sigma}_{j+1}}^{-} + h.c )
\eeq 
When $J_2 $ is real then this Hamiltonian reduces to the XY model.
Where ${{\sigma}_j}^{z} = |b_j > <b_j | ~-~ |a_j > <a_j | $,
${{\sigma}_j}^{+} = |b_j > <a_j | $, ${{\sigma}_j}^{-} = |a_j > <b_j | $
\bea
H_{xy} & = & \sum_{i=1}^{N} ( B  {{\sigma}_i}^{z}~+~ {J_1} ( {{\sigma}_i}^{x}
{{\sigma}_{i+1} }^{x}  + {{\sigma}_i}^{y}
{{\sigma}_{i+1} }^{y} ) \non\\
& & + {J_2} ( {{\sigma}_i}^{x}
{{\sigma}_{i+1} }^{x} - {{\sigma}_i}^{y}
{{\sigma}_{i+1} }^{y} ) ) \non\\    
& & =  \sum_{i=1}^{N} B ( {{\sigma}_i}^{z}~+~{J_x} {{\sigma}_i}^{x}
{{\sigma}_{i+1}}^{x} ~+~ {J_y} {{\sigma}_i}^{y}
{{\sigma}_{i+1}}^{y}) .
\eea
with ${J_x} = (J_1 + J_2 ) $ and ${J_y} = (J_1 - J_2 ) $.\\
Here we present the analytical expression of $B$, $J_1 $ and $J_2 $
in terms of different physical parameters of the system. We follow the
references \cite{hart1,james}.
\beq
B = \frac{\delta_1}{2} - \beta
\eeq
\bea
\beta  & = &  \frac{1}{2} [\frac{{|{\Omega_b}|}^2 }{4 {\Delta}_b }
({\Delta}_b - \frac{{|{\Omega_b}|}^2 }{4 {\Delta}_b } - \non\\
& & \frac{{|{\Omega_b}|}^2 }{4 ( {\Delta}_a  - {\Delta}_b )} - {\gamma_b} {g_b}^2
- {\gamma_1} {g_a}^2 + {\gamma_1}^2 \frac{{g_a}^4 }{{\Delta_b}} - (a \leftrightarrow b)]
\eea
\beq
{J_1} = \frac{\gamma_2}{4} ( \frac{{|{\Omega_a}|}^2 {g_b}^2 }{{ {\Delta}_a }^2 }
 +  \frac{{|{\Omega_b}|}^2 {g_a}^2 }{{ {\Delta}_b }^2 } )
\eeq
\beq
{J_2} = \frac{\gamma_2}{2} ( \frac{{\Omega_a} {\Omega_b} g_a g_b }{{\Delta}_a {\Delta_b} }
 ).
\eeq
where
$ \gamma_{a,b} = \frac{1}{N} \sum_{k} \frac{1}{ {\omega}_{a,b} - {\omega}_k } $
$ \gamma_{1} = \frac{1}{N} \sum_{k} \frac{1}{ ( {\omega}_{a}+  {\omega}_{b})/2 - {\omega}_k } $ and
$ \gamma_{2} = \frac{1}{N} \sum_{k} \frac{e^{ik} }{ ( {\omega}_{a}+  {\omega}_{b})/2 - {\omega}_k } $ \\
${\delta_1} = {\omega}_{ab} - ({\omega}_a - {\omega}_b )/2 $,
${\Delta}_a = {\omega}_e - {\omega}_a$.
${\Delta}_b = {\omega}_e - {\omega}_a -({\omega}_{ab} - {\delta_1})$. \\
${{\delta}_a}^{k} = {\omega}_e - {\omega}_k $,
${{\delta}_b}^{k} = {\omega}_e - {\omega}_k  -({\omega}_{ab} - {\delta_1}) $,\\
$g_a$ and $g_b$ are the couplings of respective transition to the cavity mode,
${\Omega}_a $ and ${\Omega}_b$ are the Rabi frequency of laser with frequency
$\omega_a $ and $\omega_b $. \\
Here we discuss very briefly about an effective $z$-component of 
interactions 
($ {{\sigma}_i}^{z} {{\sigma}_{i+1}}^{z}$) in such a system. The authors of 
Ref.\cite{hart1,hart2}
have proposed the same atomic level configuration but having only one
laser of frequency ${\omega}$ that mediates the atom-atom coupling through
virtual photons. Another laser field with frequency $\nu $ is used to
tune the effective magnetic field. 
In this case the Hamiltonian ${H_{AC}} $ changes but the Hamiltonians $H_A $
and $H_C $ are the same. 
\bea
{H_{AC}}&=& \sum_{j=1}^{N} [ (\frac{{\Omega}}{2} e^{-i {{\omega}}t} +
\frac{{\Lambda}}{2} e^{-i {{\nu}_a}t} 
{g_a} {a_j}) |e_j > < a_j | + h.c] \non\\
& & + [a \leftrightarrow b ] .
\eea
Here, ${\Omega}_a $ and ${\Omega}_b $ are the Rabi frequencies of the
driving laser with frequency ${\omega}$  on transition $|a > \rightarrow |e> $ 
, $|b > \rightarrow |e> $, whereas ${\Lambda}_a $ and ${\Lambda}_b $ are the 
driving laser with frequency ${\nu}$  on transition $|a > \rightarrow |e> $
, $|b > \rightarrow |e> $. One can eliminate adiabatically the excited atomic 
levels and
photons by considering the interaction picture with respect to 
$ H_0 = H_A ~+~H_C $ [6,7]. They have considered the detuning parameter in such
a way that the Raman transitions between two level are suppressed and also
chosen the parameter in such a way that the dominant two-photon processes are
those that involve one laser photon and one cavity photon but the atom
makes no transition between levels a and b. Whenever two atoms exchange a 
virtual photon both of them experience a Stark shift and play the
role of an effective $ {{\sigma}^{z}}{{\sigma}^{z}} $ interaction 
\cite{hart1,hart2,suj2}. Then
the effective Hamiltonian reduces to 
\beq
{H_{zz}}~=~\sum_{j=1}^{N} ( {B_z} {{\sigma}_j}^{z} ~+~ {J_z}
{{\sigma}_j}^{z}{{\sigma}_{j+1}}^{z} )
\eeq
These two parameters can be tuned independently by varying the laser frequencies.
Finally, they have obtained an effective model by combining Hamiltonians $H_{xy} $ and
$H_{zz} $ by using Suzuki-Trotter formalism. The effective Hamiltonian
simulated by this procedure is
\beq
H_{spin} ~=~\sum_{j=1}^{N} ( B_{tot} {{\sigma}_j}^{z} ~+~ 
\sum_{{\alpha}=x,y,z} J_{\alpha} {{\sigma}_j}^{\alpha} {{\sigma}_{j+1}}^{\alpha})
\eeq 
where $ B_{tot} = B + {B_z} $.  It has been shown in Ref. \cite{hart2} 
that $J_y $ is less than
$J_x $. From the analytical expressions of
$J_x $ and $ J_y $, it is clear that the magnitudes of ${J_1}$ and $J_2 $ 
are different. 
The result of numerical simulations trigger us also to define a model 
to study the quantum phases of this system. In the next section, we present 
the RG study of this model Hamiltonian to extract quantum phases and
transitions between them.\\
Here we present the analytical expression of $B_{tot}$, $J_z $ 
in terms of different physical parameters of the system. We follow the
references \cite{hart1,james,suj2}.\\
$ {J_z}= {\gamma}_2 {| \frac{ {{\Omega}_b}^{*} g_b }{4 \Delta_b}
- \frac{ {{\Omega}_a}^{*} g_a }{4 \Delta_a} |}^2 $ \\

$ B_{tot} = -\frac{1}{2} [ \frac{ {|{\Lambda}_b |}^2 }{16 {\tilde{\Delta_b}}^2} 
( 4 \tilde{\Delta_b } -  \frac{{|{\Lambda_a}|}^2 }{4 ( {\tilde {\Delta}_a}  - 
{\tilde {\Delta}_b} )} -  \frac{{|{\Lambda_b}|}^2 }{{\tilde {\Delta}_b } } 
- {\beta_2 }) - {\beta_3}]. $ \\

$ {\beta_2} = \sum_{j= a,b}  \frac{{|{\Omega_j}|}^2 }{4 ( {\Delta}_j  - \tilde{{\Delta}_b} )}.
4 \tilde{\gamma_{jb}} {g_j}^2  $ \\

\bea
\beta_3  & = &   [\frac{{|{\Omega_b}|}^2 }{16 { {\Delta}_b}^2 }
(4 {\Delta}_b - \frac{{|{\Omega_a}|}^2 }{4 {\Delta}_b } - \non\\
& & \frac{{|{\Omega_b}|}^2 }{4 ( {\Delta}_a  - {\Delta}_b )} - \frac{{|{\Omega_b}|}^2}{\Delta_b} 
- \sum_{j= a,b}  \frac{{|{\Lambda_j}|}^2 }{4 ( {\Delta}_j  - \tilde{{\Delta}_b} )}. 
4 {\gamma_{jb}} {g_j}^2 )  + {\gamma_{bb}}^2 \frac{{g_b}^4 }{{\Delta_b}} - (a \leftrightarrow b)]
\eea

Here ${\gamma}_1 = \frac{1}{N} \sum_{k} \frac{1}{\omega - {\omega}_k }$,
${\gamma}_2 = \frac{1}{N} \sum_{k} \frac{e^{ik}}{\omega - {\omega}_k }$,
$ {\gamma}_{aa}  = {\gamma}_{bb} = \frac{1}{N} \sum_k \frac{1}{\omega - \omega_k} $.\\

$ {\gamma}_{ab} = {\gamma}_{ba} = \frac{1}{N} \sum_k \frac{1}{\omega \pm \omega_{ab} - {\omega}_k } $
$\tilde{ {\gamma}}_{ab} = \tilde{{\gamma}}_{ba}= 
\frac{1}{N} \sum_k \frac{1}{\nu \pm \omega_{ab} - {\omega}_k } $\\

$ \tilde{{\gamma}}_{aa} = \tilde{{\gamma}}_{bb}= \frac{1}{N} \sum_k \frac{1}{\nu - \omega_k} $.

{\bf Analytical Derivation and Analysis of RG Equations:}

To study the different quantum phases of the system described by
the Hamiltonian (Eq. 13), we express this Hamiltonian in more explicit way,
\bea
H_{2} ~=~ \sum_n ~[ & & (1+a) ~S_n^x S_{n+1}^x ~+~ (1-a)~ S_n^y S_{n+1}^y \non \\
& & +~ \Delta ~S_n^z S_{n+1}^z ~+~ h ~S_n^z ~]~,
\label{ham2}
\eea
where $S_n^{\alpha}$ are the spin-1/2 operators.
We assume that the $XY$ anisotropy $a$ and the $zz$ coupling $\Delta$
satisfy the condition $-1 \le \Delta \le 1$, and $ 0 < a \leq 1 $
and magnetic field strength is $h \ge 0$.
The parameters correspondence
between the micro cavities and spin chain are the following, 
$h \sim B_{tot}$, ${\Delta = J_z}$, ${J_1 =1 }$ and ${J_2} =a $.
The $XY$ anisotropy breaks the in plane rotational symmetry. The study
of the quantum phases from the perspective of 
quantum spin system and magnetism is not 
entirely a new one \cite{suj,ric,zamo}. Here our main aim is to study the quantum
phases of microcavities array through the RG analysis of this
model Hamiltonian.  
\\
Spin operators can be recast in terms of spinless fermions through
Jordan-Wigner
transformation and then finally one can express the spinless fermions
in terms of bosonic fields \cite{gia}.
We recast the spinless
fermions operators in terms of field operators by this relation.
$ {\psi}(x)~=~~[e^{i k_F x} ~ {\psi}_{R}(x)~+~e^{-i k_F x} ~ {\psi}_{L}(x)] $
, where ${\psi}_{R} (x)$ and ${\psi}_{L}(x) $ describe the second-quantized
fields of right- and
the left-moving fermions respectively,
and $k_F$ is the Fermi wave vector.
We express the fermionic fields in terms of bosonic
field by the relation
$ {{\psi}_{r}} (x)~=~~\frac{U_r}{\sqrt{2 \pi \alpha}}~
~e^{-i ~(r \phi (x)~-~ \theta (x))},$
where $r$ denotes the chirality of the fermionic fields,
right (1) or left movers (-1).
The operators $U_r$ is the Klein factor to preserve the anti-commutivity of
fermions. 
$\phi$ field corresponds to the
quantum fluctuations (bosonic) of spin and $\theta$ is the dual field of $\phi$.
They are
related by the relations
$ {\phi}_{R}~=~~ \theta ~-~ \phi$ and  $ {\phi}_{L}~=~~ \theta ~+~ \phi$.
Hamiltonian 
$H_0  = \frac{v}{2} ~\int ~dx ~[~ (\partial_x \theta)^2 ~+~ (\partial_x
\phi)^2 ~]  $ is non-interacting part of $H_{XYZ}$.
Here $v$ is the velocity of the low-energy excitations. It is one of the Luttinger
liquid parameters and the other is $K$, which is related to $\Delta$
by \cite{gia,ric}
\beq
K ~=~ \frac{\pi}{\pi + 2 \sin^{-1} (\Delta)} ~.
\label{kd}
\eeq
where $K$ takes the values 1 and 1/2 for $\Delta =0$ (free field), and $\Delta =1$
(isotropic anti-ferromagnet), respectively.
The relation between $K$ and $\Delta$ is not preserved under the
renormalization, so this relation is only correct for the initial
Hamiltonian.
The analytical form of the spin operators
in terms of the bosonic fields are:
$ S_n^x ~=~ [~ c_2 \cos (2 {\sqrt {\pi K}} \phi) ~+~ (-1)^n c_3 ~]~
\cos ({\sqrt {\frac{\pi}{K}}} \theta )$; 
$S_n^y ~=~ -[~ c_2 \cos (2 {\sqrt {\pi K}} \phi) ~+~ (-1)^n c_3 ~]~
\sin ({\sqrt {\frac{\pi}{K}}} \theta )$, and 
$ S_n^z ~=~ {\sqrt {\frac{\pi}{K}}} ~\partial_x \phi ~+~ (-1)^n c_1
\cos (2 {\sqrt {\pi K}} \phi ) ~$
where $c_i$'s are constants as given in Ref. \cite{zamo}.
The Hamiltonian $H_{2}$ in terms of bosonic fields is the
following,
\bea
H_{2} &=& H_0 + \frac{a}{2 \pi \alpha} \int \cos (2 {\sqrt {\frac{\pi}{K}}}
\theta (x) ) dx  \non \\ 
& & + \frac{\Delta}{{2 \pi \alpha}^2} \int \cos(4 {\sqrt{\pi K}} \phi (x)) dx 
 + \frac{h \sqrt{K }}{\pi \alpha} \int {{\partial}_x } {\phi (x) } dx 
\label{ops1}
\eea
One can also write the above Hamiltonian in the following form
\bea
H_{2} &=& H_{01} + \frac{a}{2 \pi \alpha} \int \cos (2 {\sqrt {{\pi}}}
\theta (x) ) dx  \non \\ 
& & + \frac{\Delta}{{2 \pi \alpha}^2} \int \cos(4 {\sqrt{\pi }} \phi (x)) dx 
 + \frac{h }{\pi \alpha} \int {{\partial}_x } {\phi (x) } dx 
\label{ops1}
\eea
Where $H_{01} $,
\bea
H_{01} = \frac{1}{2 \pi} \int dx [u K {({\nabla \theta (x)})}^2 
+ (u/K) {({\nabla \phi (x)})}^2 ]
\eea 
One can get the $H_{XY}$ Hamiltonian by simply putting $\Delta =0$
in the above Hamiltonian.
In this derivation, different powers of
coefficients $c_i$ have been absorbed
in the definition of $a, h$ and $\Delta$.
The integration of the oscillatory terms in the Hamiltonian yield
negligible small contributions and the origin of the oscillatory terms  
occur due the spin operators.
So it's a reasonably good approximation to keep only the non-oscillatory
terms in the Hamiltonian.
The Gaussian scaling dimension of these
coupling terms, $a$ and $\Delta$ are $1/K$ and $4K$ respectively.
The third term ($ \Delta $) of the Hamiltonian tends to order 
the system into density wave phase 
, whereas the second term 
($ a $) of the Hamiltonian favors the staggered order in 
the $XY$ plane. Two sine-Gordon coupling terms are from two dual
fields. Therefore, the model Hamiltonian consists of two competing
interactions. 
This Hamiltonian contains two 
strongly relevant and mutually nonlocal perturbation over
the Gaussian (critical) theory.
In such a situation the strong coupling fixed point is usually
determined by the most relevant perturbation whose amplitude
grows up according to its Gaussian scaling dimensions and
it is not much affected by the less relevant coupling terms.
However, this is not the general rule if the two operators
exclude each other, i.e., if the field configurations which
minimize one perturbation term do not minimize the other.
In this case interplay between the two competing relevant
operators can produce a novel quantum phase transition through
a critical point or a critical line. Therefore, we would like to
study the RG equation to interpret the quantum phases of the system.
In the RG theory, we not only able to predict the weak coupling
limit but also the the intermediate values of the coupling. The RG theory is
a perturbative theory and it ceases to be valid when the coupling
constant $g(l) \sim 1$. \\
We now study how the parameters $a$, $\Delta$ and $K$ flow under RG.
The operators in Eq. (17) are related to each other
through the
operator product expansion. So the RG equations for their coefficients
therefore are coupled to each other.
We use operator product expansion to derive
these RG equations which is independent of boundary condition
\cite{cardy}. 
In our derivation, we 
consider two operators,
$ X_1 = e^{(i a_1 \phi + i b_1 \theta)}$ and $ X_2 = e^{(i a_2 \phi + i b_2 \theta)}$.
In the RG procedure, one can write these two field operators as a sum of
fast and slow mode fields. In the fast field, the momentum range is 
$ \Lambda e^{-dl} < K < \Lambda $ and for the slow field 
$ K < \Lambda e^{-dl} $,
where $\Lambda $ is the momentum cut-off, $dl $ is the change in the logarithmic
scale. The next step is the integration of the fast field for the operators $X_1 $
and $X_2 $, it yields a third operator at the same space time point,
$ X_3 = e^{i (a_1 + a_2) \phi + i (b_1 + b_2) \theta)}$. The prefactor of $X_3 $
can be found by the relation, $ {X_1 }{X_2} \sim e^{-(a_1  a_2 + b_1 b_2 )} \frac{dl}{2 \pi} X_3 $.
Our Hamiltonian consists of two operators, if we consider $l_1 $ and $l_2 $ as the coefficient of
the operators $X_1 $ and $X_2 $ respectively. Then the RG expressions for $\frac{d X_3}{dl}$
contains the term $ (a_1 a_2 + b_1 b_2 ) \frac{l_1 l_2}{2 \pi} $. This is the procedure
to derive these RG equations.\\
In the RG process, one can write 
RG equations themselves are established
in a perturbative expansion in coupling constant ($g(l)$). They
cease to be valid beyond a certain length scale, where
$g(l) \sim 1$ \cite{gia}.
The RG equations for the coefficients of Hamiltonian $H_{XYZ}$ are
\bea
\frac{da}{dl} ~&=&~ (2 -\frac{1}{K}) a ,\non \\
\frac{d{\Delta}}{dl} ~&=&~ (2 - 4K) {\Delta}  \non \\
\frac{dK}{dl} ~&=&~ \frac{a^2}{4} ~-~ K^2 {\Delta}^2 ~,
\label{rg2}
\eea
We have followed Ref.\cite{suj} during the derivation of these RG equations.
These RG equations have trivial (${a^*}= 0 = {{\Delta}^*}$)
fixed points for any arbitrary $K$. 
Apart from that these RG equations have also two non-trivial
fixed lines, $a = \Delta $ and $a = - {{\Delta} }$
for $ K =1/2 $.
The above RG equations show that there is no duality in flow diagram. 
Here we mean, duality, that if $\theta$ and
$\phi $ interchange $ \theta \leftrightarrow \phi$, $ K \leftrightarrow K^{-1}$ and
$ \Delta \leftrightarrow a$ will not produce the same set of RG equations.\\
For $ K > 1/2 $, the sine-Gordon coupling term correspond to the
anisotropic exchange coupling become relevant and  
the system flows to the photonic superfluid phase. For $ K < 1/2 $,
the sine-Gordon coupling term correspond to the z-component exchange
interaction become relevant and the system flowing off to the
photon blocked induced insulating phase.\\ 
Here we explain the physical 
significance of different quantum phases 
of the atom-cavity system 
what we find in our study.\\  
(1). When both the anisotropic exchange interaction and the z-component of 
exchange interaction is absent then there are no sine-Gordon coupling
terms in the Hamiltonian. Then the system is in the mass less Luttinger 
liquid phase, i.e., the system is in the photonic metallic state.
The other source of photonic metallic state is that when 
the system shows the BKT transition which we will discuss in the 
next section. 
\\
(2). The photonic insulator state of the atom-cavity array   
system corresponds to the insulating state of the system where there is no
transmission of photon between the microcavities in the array due to the
interaction between photons. The appearance of this phase has already discussed
in the analysis of three RG equations. This phase will occurs
when the system shows the BKT transition, which we will discuss in the next
section. \\
(3). Photonic superfluid state of the atom-cavity array
system corresponds to the gapless excitations of the system where the
photon transmit from one cavity to the other without any blocking 
. For this one dimensional
cavity QED system where there is no order parameter, one can only discriminate 
between the photonic metallic state and photonic superfluid state by
only finding the difference of fluctuation in photon number in every
sites of the array. This photonic superfluid phase corresponds to
dissipitionless flow of photon in the system. 
We have already discussed about the appearance of photonic superfluid phase
from the analysis of three RG equations. The other source of the
appearance of photonic superfluid phase is the BKT transition which we will
discuss in the next section. \\

{\bf {\large Berezinskii-Kosterlitz-Transition Physics in Cavity QED System. 
}} \\
The physics of BKT transition has found in different one dimensional and
two dimensional (classical system) and it has discussed extensively in
different context in the Ref. \cite{subir,gia,druf}.\\
Before we start to discuss the appearance of BKT transition in our
system, we would like to discuss very briefly why it is necessary
to study the BKT transition. Here we study two different situations
of our model Hamiltonian. For the first case the exchange anisotropy
is absent ($J_2 =0 $) and for the second case z-component of exchange
interaction is absent ($J_z =0 $). For both of these cases only
one of the sine-Gordon coupling term is present, therefore, there is
no competition between the two mutually non local perturbation. Therefore
one can think that there is no need to study the RG to extract the
quantum phases and phase boundaries. But we still apply RG method for
the following reason. Each of these Hamiltonians consist of two part,
the first one ($H_{01}$ ) is the non-interacting where the $\phi$ and
$\theta$ fields show the quadratic fluctuations and the other part 
of these Hamiltonians are the sine-Gordon coupling terms which  
of either $\theta$ or $\phi $ fields. The sine-Gordon coupling term
lock the field either $\theta$ or $\phi$ in the minima of the
potential well. Therefore the system has a competition between
the quadratic part of the Hamiltonian and the sine-Gordon
coupling term and this competition will govern the low energy
physics of these Hamiltonians in different limit of the
system. The RG process (BKT transition)
will predict the quantum phases of these system correctly.\\ 
It is very clear from the analytical expression of $ {J_1}, J_2 $
and $ J_z $ that one can control these parameters in the 
laboratory. The quantum state of engineering of cavity QED system is in the
state of art due to the rapid 
technological development of this
system [1]. In this study we consider the situations where the $J_2 $ 
is absent, i.e., the system with a single Rabi frequency oscillation.
We also consider the situation where the z-component of exchange interaction
is absent. These two situations lead to the two different set of RG equations
which show BKT transition.\\
For the first case, there is no anisotropic exchange
coupling. In this situation, the three RG equations of the previous section reduce
to two RG equations, which are the following  
\bea
\frac{d{\Delta}}{dl} ~&=&~ (2 - 4K) {\Delta}  \non \\
\frac{dK}{dl} ~&=&~ -~ K^2 {\Delta}^2 ~,
\label{rg3}
\eea
Now we express these RG equations in the form of BKT transition form. As we
understand from the RG equations that the transition occurs
at $K=1/2 $. 
To study the flow of the RG equation around this transition point,
we recast the RG equation in suitable form.\\ 
Here we follow the following transformation,
$ K = 1/2 + \frac{y_{||}}{4} $ and $ \Delta \rightarrow \Delta/2 $.\\
The above equation reduce to the standard BKT equation.\\
\bea
\frac{d y_{||} }{dl} ~&=&~ -  {\Delta}^2   \non \\
\frac{d \Delta }{dl} ~&=&~ -~ y_{||} {\Delta} ~,
\label{rg4}
\eea
In our case, $ \Delta \frac{d \Delta}{dl} = y_{||} \frac{d y_{||}}{dl} $.
Therefore $ X^2 = {y_{||}}^2 - {\Delta}^2 $ is a constant of motion. 
Here
we discuss the relevant physics and the quantum phase transition 
between the photonic metallic state and photonic Coulomb blocked 
state based
on these equations. Here we consider the following situations based on these
equations:\\
(1) When $X >0 $ and $ y_{||} >0 $, the sine-Gordon coupling term
corresponding to $\Delta$ term is irrelevant, the fixed point,
$ {\Delta}^{*} =0 $ and $ {y_{||}}^{*} = X $, close to this
fixed point, we can write the RG equations as
 $ \frac{d \Delta}{dl} = 2 (1 -2 K^* ) {\Delta (l)} $ and $ \frac{dy_{||}}{dl} = 0 $.
Using the flow equation and constant of motion for $ {y_{||}} > {\Delta} $.
One can write the solution of coupling terms as
\beq
y_{||} (l) = \frac{X}{tanh (Xl + atanh(\frac{X}{{y_{||}}^0 }) }
\eeq
\beq
\Delta (l) = \frac{X}{sin(Xl + atanh(\frac{X}{{y_{||}}^0 }) }
\eeq
The condition at the line of seperatix is
$ {y_{||}} (l) = {\Delta (l) }= \frac{ {\Delta}^0 }{1 + {\Delta}^0 l } $.
There is no mass gap excitation in the elementary excitation of the system, 
i.e, there is no photon blocked induced insulating phase. 
Hence the system is in the photonic
metallic state.\\
To the best of our knowledge for the first time in the literature we predict
the existence of photonic metallic state for the array of Cavity QED system.
Here we would like to present the basic origin of photonic-metallic state
explicitly.\\
If we do the Jordan-Wigner transformation of our model Hamiltonian 
as we present in Ref. 22.
It is very clear from the 
Eq. 32 of Ref. 22 that the first term represent the photon hopping term across the
lattice of the cavity QED array. The second and third term of the Hamiltonian 
represent the photonic pair correlation and photonic density wave respectively in the system.
When the second and third term in the Hamiltonian either are absent or irrelevant
in the RG sense at that situation the system shows the photonic metallic state.\\ 

(2) When $ {\Delta} > y_{||} $. The RG equation for the coupling $ \Delta (l)$ flowing
off to the strong coupling regime. A perturbative expansion in $\Delta $ cease to be 
valid beyond a certain length scale for which $ \Delta (l) \sim 1$. The analysis 
of the RG equation is not valid beyond this length scale.\\
The analytical relation between the coupling constant is
\beq
arctan ( \frac{ {{y}_{||}}^{0} }{ \sqrt{ {{\Delta}_0}^2 - {{y_{||}}^0}^2 } } )
- arctan ( \frac{ {{y}_{||}} }{ \sqrt{ {{\Delta}_0}^2 - {{y_{||}}^0}^2} } )
= \sqrt{ {{\Delta}_0}^2 - {{y_{||}}^0}^2 }.
\eeq
In this limit, the RG flowing off to the strong coupling phase. Here we
discuss the relevant physics of the phase. 
The sine-Gordon coupling term is
$$ \frac{\Delta_1  u}{ 2 \pi {\alpha}^2 } \int dx cos ( 4 \phi (x) ).$$
where $\Delta_1 = \frac{\Delta}{\pi u} $.
The $\phi (x) $ field locks into one of the minima of the cosine potential. 
Now we expand the potential for large $\Delta_1 $ in the spirit
of usual RG method \cite{subir,gia,druf}. We can write the effective
Hamiltonian near to the minima\\
$$ H = H_0 + \frac{4 \Delta_1 u }{ \pi {\alpha}^2 } {\phi}^2 (x) $$.
The total action of the Hamiltonian of the system can be written as
\beq
S = \frac{1}{2 \pi K} \frac{1}{\beta \Omega} \sum_{k, {\omega}_n }
[ \frac{ {{\omega}_n}^2 }{u} + u k^2 + \frac{8 K \Delta_1 u}{{\alpha}^2 }]
{\phi}^* (k, \omega_n ) {\phi} (k, \omega_n )
\eeq 
The excitations of the system which cost a finite energy even at $k =0 $.
In this situation the field, $\phi (x) $ is massive. 
The system posses phononic type mode.   
This phononic mode is the small oscillation
of the field $ {\phi} (x) $ around the minima of the cosine potential.\\
Suppose we consider the RG equation up to the point where ${\Delta_1}(l) \sim 1$.
This excitations gap in the spectrum has the dimension of an energy and the
renormalization relation is ${\Omega}_{M} (l) = e^{l} {\Omega}_{M} (l=0) $,
for the case, ${\Delta_1} (l^* ) \sim 1 $. We can write the 
expression for the gap following the action as,
$ {\Omega}_{M} (l^* ) \sim  \sqrt{ {\Delta_1 (l^* )}} u /{\alpha}$. 
The true gap of the system is 
$ {\Omega}_{M} (l=0) \sim e^{-l^*} {\Omega}_{M} (l^* ) $, 
${\Omega}_{M} (l^* ) = \frac{{u}}{\alpha} $.
Thus at this phase of this system, the explicit dependence of
the Luttinger liquid parameter is absent. This gap of the system is of the order of bandwidth
of the cavity QED system.\\

(3) Now we consider the case, when ${\Delta_1} << | y_{||} |$. The system is in the
deep massive phase. In this phase, one can write the RG equation as 
$ \frac{d K(l)}{d l} =0$ and $ \frac{d \Delta_1 (l)}{dl} = 2 (1 - 2K) {\Delta_1} (l) $.
For this RG equation, we get ${\Delta_1} (l) = {\Delta_1} (0) e^{2 (1 - 2K)l}$ from this
equation $ e^{- l^* } = {\Delta (0)}^{1/2(1- 2K)} $. The true gap of the 
system is 
${\Omega}_2 (l=0) \sim e^{-l^*} {\Omega}_2 (l^* ) $, 
$\frac{\Omega_2 (l=0) }{\Omega_2 (l^* ) } \sim {\Delta_1 (0)}^{+1/2(1-2K)} $. This gap is
the power law dependence of the bare $\Delta_1$. This gap varies with $K $, i.e.,
the gap of the system is now varying with the interaction of the system. It is very
clear from the analytical expression that the gap gets 
smaller and tends to zero at $K=1/2$, this prediction
is consistent with the physical scenario that at $K= 1/2$ the system shows the phase
transition.\\ 

(4) Close to the transition point, Eq.(25) reduces to 
$ \sqrt{ {{\Delta_1}_0}^2 - {(y_{||})}^2 } {l^* } = \pi$.
So the square root term goes to zero at the transition point and therefore 
${y_{||}}^{0}/{\sqrt{ {{\Delta_1}_0}^2 - {y_{||}}^2 }} \rightarrow \infty$ and 
the gap 
${\Omega} (l =0)  \sim {\Omega}({l^* }) e^{- \frac{ \pi}
{ \sqrt{ {{\Delta_1}_0}^2 - {y_{||}}^2 } } }  $. 
As one approach to the transition point, 
such as $ {\Delta_1}_0 \rightarrow |{y_{||}}| $. The gap is exponentially 
small in the square root of the distance to the transition. 
\\
It is therefore clear from our RG analysis that the system is in the massive phase
for the two limits. For the first case when $ {\Delta} >> {y_{||}} $, the excitation 
gap of the system is of the order of the bandwidth of the system and the effect of
photonic strong correlation is not explicit. The other limit of gapped state, i.e,
when $ \Delta << |y_{||}| $. The excitation gap of the system has a power law
dependence which vary with $K$.\\  
It is very clear from the analytical expression of $J_z$ that one can tune it to zero
by adjusting the Rabi frequencies oscillation of the system.
In this situation, there is no z-component of intercavity exchange interaction.
Therefore the three RG equations are reduce to two RG equations.\\ 
\bea
\frac{da}{dl} ~&=&~ (2 - 1/K) a  \non \\
\frac{dlnK}{dl} ~&=&~ ~ {a^2 }/{4K}  ~,
\label{rg3}
\eea
If we do the following transformation $\tilde{K} = 1/{2 K}$ and
$\tilde{a} = a/2 $ and after that we follow another transformations
$\tilde{K} = 1 + {y_{||}}/2 $ the above equations reduce to 
\bea
\frac{d y_{||} }{dl} ~&=&~ -  {\tilde a}^2   \non \\
\frac{d \tilde a }{dl} ~&=&~ -~ y_{||} {\tilde a} ~,
\label{rg4}
\eea
The mathematical structure of these equations are the same as that of 
Eq. 22. Therefore the mathematical analysis 
for the different
limits are the same but massive phase and deep massive phase for these
situations are the photonic superfluid phase. 
\\ 
The sine-Gordon coupling term is
$$ \frac{a_2  u}{ 2 \pi {\alpha} } \int dx cos ( 2 \theta (x) ).$$
where $a_2 = \frac{a}{u} $.
The $\theta (x) $ field locks into one of the minima of the cosine potential. 
Now we expand the potential for large $a_2 $. We can write the effective
Hamiltonian near to the minima\\
$$ H = H_0 + \frac{a_2 u }{ { \pi \alpha} } {\theta}^2 (x). $$
The total action of the Hamiltonian of the system can be written as
\beq
S = \frac{1}{2 \pi K} \frac{1}{\beta \Omega} \sum_{k, {\omega}_n }
[ \frac{ {{\omega}_n}^2 }{u} + u k^2 + \frac{2 K a_2 u}{{\alpha} }]
{\theta}^* (k, \omega_n ) {\theta} (k, \omega_n )
\eeq 
The excitations of the system which cost a finite energy even at $k =0 $.
In this situation the field, $\theta (x) $ is massive. 
The true gap of the system is 
$ {\Omega}_{C} (l=0) \sim e^{-l^*} {\Omega}_{C} (l^* ) $, 
${\Omega}_{C} (l^* ) = (\frac{{u}}{\alpha}) \sqrt{\alpha} $.
Thus at this phase of this system, the explicit dependence of
the Luttinger liquid parameter is absent. This gap of the system is 
much less than the excitation gap of the photon blocked state due to the
presence of the extra factor $\sqrt{\alpha}$.\\
In the deep massive phase,   
$\frac{\Omega_C (l=0) }{\Omega_C (l^* ) } \sim {a_2 (0)}^{+1/(2-1/K)} $. This gap is
the power law dependence of the bare $a_2$. This gap varies with $K $, i.e.,
the gap of the system is now varying with the interaction of the system. It is very
clear from the analytical expression that the gap gets 
smaller and tends to zero at $K=1/2$ and 
this prediction
is consistent with the physical scenario that at $K= 1/2$ the system shows the phase
transition from photonic metallic state to photonic superfluid phase.\\ 
Now we consider the effect of magnetic field ( ${\delta}_1 $ term) 
on the quantum phases and phase boundaries of this cavity QED arrays.\\
We do the
following transformation to eliminate the magnetic field term from the Hamiltonian.
We substitute 
$
 4 \pi \sqrt{K} \phi  \rightarrow  4 \pi \sqrt{K} \phi 
+ 4 \pi  \delta x 
$,  
where $ \delta = \sqrt{K} h $. This introduce the spatial oscillation 
in the $\Delta $ term. When the coupling $a$ is absent and the 
applied detuning field is larger than the photonic Mott gap, the
system drives to the gapless photonic metallic
state. This is nothing but the well
known Porkovosky-Talapov model which shows the 
commensurate to incommensurate transition \cite{gia,suj}.
When $ \delta {a_1} >> 1$ (where $a_1$ is the lattice spacing), 
the $cos( 4 \pi \sqrt{K} (\phi + \delta x) ) $ term is quickly
oscillating and averages out zero. Thus the system reflects the competition between the
$\Delta$ and $h$. As a results of it, the RG flow for $\Delta $ has to be cutoff
when $4 \pi \sqrt{K} \delta (l) a_1 \sim 1 $. To the lowest order in $\Delta, a $
and $\delta$, the RG flow equation is 
\beq
\frac{d \delta}{dl} = \delta
\eeq
As a result of this RG equation, the system shows two different
response, either the flow ${\Delta}_1 $ and $\delta $
or $a$ and $\delta $ flows of to the strong coupling phase.\\
When $\Delta_1 (l)$ reaches strong coupling before $ \delta a_1 $
become order one. The phase boundary is the same as we predict without 
magnetic field. The condition when $\Delta_1  (l^*) \sim 1$ define a new
length scale $ l^* $. 
It generates a self-consistent scenario that 
$4 \pi \delta (l^* ) a_1 << {{\Delta_1} (0) }^{1/(2 - 4 K) }$.  
In the other limit when 
the RG flow of $a$ flows off to the strong coupling phase 
does not affected by the magnetic field term and thus the system is in the
photonic superfluid phase.\\  
{\bf Conclusions:}  
We have presented three sets of RG equations for the different physical
situations for the Cavity QED system. 
We have predicted two different BKT transitions for the different
physical situations.
We have predicted the photonic
superfluid phase, photonic metallic phase and photon blocked induced 
insulating phase. To the best of our knowledge, this explicit quantum
phase analysis and their behavior based on the RG study 
for the microcavity array system is absent
in the previous literature. 
\centerline{\bf Acknowledgments}
\vskip .2 true cm
The author would like to acknowledge the series of lectures and also discussions
of Prof. S. M. Girvin during the International Workshop/School on Dirac Materials
and Chandrashekar Discussion Meeting, December'2012.
The author would like to thank The Center for Condensed Matter 
Theory of the 
Physics Department
of IISc for extended facility. Finally the author would like to thank Prof. Prabir
Mukherjee for reading this manuscript critically.

\end{document}